\documentclass[iop]{emulateapj}

\usepackage{amsmath,natbib,graphicx}
\usepackage{epsf}
\usepackage{epstopdf}
\usepackage{epsfig}
\usepackage{color}
\DeclareGraphicsExtensions{.jpg,.pdf,.png,.eps,.ps}
\usepackage{scrextend}
\usepackage{hyperref}

\def\UFlorida{1}
\def\CfA{4}
\def\Illinois{2}
\def\Arizona{11}
\def\MPIfR{15}
\def\ESOGarching{5}
\def\Stanfordnb{10, 18}
\def\Stanford{10}
\def\UCLA{12}
\def\Dal{9}
\def\Diego{3}
\def\Cambridge{6}
\def\PSUnb{7, 16,17}
\def\PSU{7}
\def\KICPChicago{8}
\def\MIT{13}
\def\NRAO{14}

\begin{document}

\title{SPT0346-52: negligible AGN activity in a compact, hyper-starburst galaxy at $z$ = 5.7}

%\shorttitle{}

\newcommand{\flux}{ergs cm$^{-2}$ s$^{-1}$}
\newcommand{\lum}{ergs s$^{-1}$}

%%%%%%%%%%%%%%%%%%%%%%%%%%%
% AUTHORS AND AFFILIATIONS
%%%%%%%%%%%%%%%%%%%%%%%%%%%

\shortauthors{J. Ma, et al.}

\author{
Jingzhe Ma$^{\UFlorida}$, 
Anthony.~H.~Gonzalez$^{\UFlorida}$, 
J.~D.~Vieira$^{\Illinois}$, 
M.~Aravena$^{\Diego}$, 
M.~L.~N.~Ashby$^{\CfA}$,
M.~B\'ethermin$^{\ESOGarching}$, 
M.~S.~Bothwell$^{\Cambridge}$,
W.~N.~Brandt$^{\PSUnb}$, 
C.~de~Breuck$^{\ESOGarching}$, 
J.~E.~Carlstrom$^{\KICPChicago}$,
S.~C.~Chapman$^{\Dal}$,
B.~Gullberg$^{\ESOGarching}$,
Y.~Hezaveh$^{\Stanfordnb}$,
K.~Litke$^{\Arizona}$,
M.~Malkan$^{\UCLA}$,
D.~P.~Marrone$^{\Arizona}$, 
M.~McDonald$^{\MIT}$,
E.~J.~Murphy$^{\NRAO}$,
J.~S.~Spilker$^{\Arizona}$,
J.~Sreevani$^{\Illinois}$, 
A.~A.~Stark$^{\CfA}$,
M.~Strandet$^{\MPIfR}$,
S.~X.~Wang$^{\PSU}$
}

\altaffiltext{\UFlorida}{Department of Astronomy, University of Florida, Gainesville, FL 32611, USA; \href{mailto:jingzhema@ufl.edu}{jingzhema@ufl.edu}}
\altaffiltext{\Illinois}{Department of Astronomy and Department of Physics, University of Illinois, 1002 West Green St., Urbana, IL 61801, USA}
\altaffiltext{\Diego}{N\'ucleo de Astronom\'{\i}a, Facultad de Ingenier\'{\i}a, Universidad Diego Portales, Av. Ej\'ercito 441, Santiago, Chile}
\altaffiltext{\CfA}{Harvard-Smithsonian Center for Astrophysics, 60 Garden Street, Cambridge, MA 02138, USA}
\altaffiltext{\ESOGarching}{European Southern Observatory, Karl Schwarzschild Stra\ss e 2, 85748 Garching, Germany}
\altaffiltext{\Cambridge}{Cavendish Laboratory, University of Cambridge, JJ Thompson Ave, Cambridge CB3 0HA, UK}
\altaffiltext{\PSU}{Department of Astronomy and Astrophysics, The Pennsylvania State University, University Park, PA 16802, USA}
\altaffiltext{\KICPChicago}{Kavli Institute for Cosmological Physics, University of Chicago, 5640 South Ellis Avenue, Chicago, IL 60637, USA}
\altaffiltext{\Dal}{Dalhousie University, Halifax, Nova Scotia, Canada}
\altaffiltext{\Stanford}{Kavli Institute for Particle Astrophysics and Cosmology, Stanford University, Stanford, CA 94305, USA}
\altaffiltext{\Arizona}{Steward Observatory, University of Arizona, 933 North Cherry Avenue, Tucson, AZ 85721, USA}
\altaffiltext{\UCLA}{Department of Physics and Astronomy, University of California, Los Angeles, CA 90095-1547, USA}
\altaffiltext{\MIT}{Kavli Institute for Astrophysics and Space Research, Massachusetts Institute of Technology, 37-582C, Cambridge, MA 02139, USA}
\altaffiltext{\NRAO}{National Radio Astronomy Observatory, 520 Edgemont Road, Charlottesville, VA 22903, USA}
\altaffiltext{\MPIfR}{Max-Planck-Institut f\"{u}r Radioastronomie, Auf dem H\"{u}gel 69 D-53121 Bonn, Germany}
\altaffiltext{16}{Institute for Gravitation and the Cosmos, The Pennsylvania State University, University Park, PA 16802, USA}
\altaffiltext{17}{Department of Physics, The Pennsylvania State University, University Park, PA 16802, USA}
\altaffiltext{18}{Hubble Fellow}

%%%%%%%%%%%%%%%%%%%%%%%%%%%
% ABSTRACT
%%%%%%%%%%%%%%%%%%%%%%%%%%%

\begin{abstract}
We present {\it Chandra} ACIS-S and ATCA radio continuum observations of the strongly lensed dusty, star-forming galaxy SPT-S J034640-5204.9 (hereafter SPT0346-52) at $z$ = 5.656. This galaxy has also been observed with ALMA, {\it HST}, {\it Spitzer}, {\it Herschel},  APEX, and the VLT. Previous observations indicate that if the infrared (IR) emission is driven by star formation, then the inferred lensing-corrected star formation rate ($\sim$ 4500 $M_{\sun}$ yr$^{-1}$) and star formation rate surface density $\Sigma_{\rm SFR}$ ($\sim$ 2000 $M_{\sun} {\rm yr^{-1}} {\rm kpc^{-2}}$) are both exceptionally high. It remained unclear from the previous data, however, whether a central active galactic nucleus (AGN) contributes appreciably to the IR luminosity. The {\it Chandra} upper limit shows that SPT0346-52 is consistent with being star-formation dominated in the X-ray, and any AGN contribution to the IR emission is negligible. The ATCA radio continuum upper limits are also consistent with the FIR-to-radio correlation for star-forming galaxies with no indication of an additional AGN contribution. The observed prodigious intrinsic IR luminosity of (3.6 $\pm$ 0.3) $\times$ 10$^{13}$ $L_{\sun}$ originates almost solely from vigorous star formation activity. With an intrinsic source size of 0.61 $\pm$ 0.03 kpc, SPT0346-52 is confirmed to have one of the highest $\Sigma_{\rm SFR}$ of any known galaxy. This high $\Sigma_{\rm SFR}$, which approaches the Eddington limit for a radiation pressure supported starburst, may be explained by a combination of very high star formation efficiency and gas fraction. 

\end{abstract}
\keywords{galaxies: high-redshift}

%%%%%%%%%%%%%%%%%%%%%%%%%%%
% I INTRODUCTION
%%%%%%%%%%%%%%%%%%%%%%%%%%%
\section{Introduction}\label{sec:intro}

A population of gravitationally lensed dusty star-forming galaxies (DSFGs) has been discovered by the South Pole Telescope (SPT) survey \citep{Vieira2010} and facilitated our understanding of the stellar, gas, and dust content of the high-redshift Universe.  One of the sources stands out as the most extraordinary discovered so far in the 2500 deg$^2$ survey: SPT-S J034640-5204.9 (hereafter SPT0346-52) at $z$ = 5.656, among the highest-redshift DSFGs known. It has been the focus of a multi-wavelength observational campaign with {\it HST}, {\it Spitzer}, {\it Herschel}, the Atacama Large Millimeter/submillimeter Array (ALMA), the Atacama Pathfinder EXperiment (APEX), and the Very Large Telescope (VLT).

Our lens model obtained from ALMA 870 $\micron$ imaging shows that SPT0346-52 is magnified by the foreground lensing galaxy (at $z$ $\sim$ 1.1) a factor of 5.6 $\pm$ 0.1 with an intrinsic 870 $\micron$ flux of 19.6 $\pm$ 0.5 mJy, and has an intrinsic size of $R_{\rm eff}$ = 0.61 $\pm$ 0.03 kpc ($R_{\rm eff}$ being half-light radius; \citealt{Hezaveh2013, Spilker2016}).  Multi-band spectral energy distribution (SED) fitting gives an intrinsic infrared (IR; 8-1000${\micron}$) luminosity $L_{\rm IR}$ of (3.6 $\pm$ 0.3) $\times$ 10$^{13}$ $L_{\sun}$ and a star formation rate (SFR) of 4500 $\pm$ 1000 $M_{\sun}$ yr$^{-1}$ \citep{Ma2015}. Given its size, SPT0346-52 turns out to have one of the highest IR luminosity surface density and SFR surface density $\Sigma_{\rm SFR}$ of any known galaxy (\citealt{Hezaveh2013,Spilker2015,Spilker2016}). The central question is whether this high luminosity surface density arises solely from intense star formation, or if there is an obscured active galactic nucleus (AGN).

\begin{figure}[ht]
\centering
{\includegraphics[width=9cm, height=7.5cm]{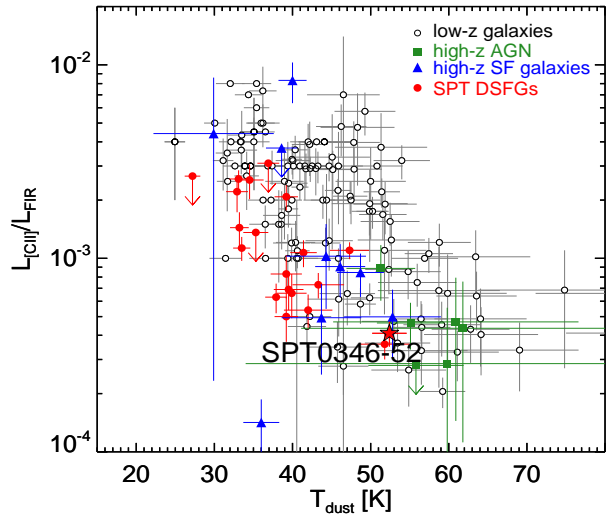}} 
\caption{$L_{\rm [CII]}/L_{\rm FIR}$ vs. $T_{\rm dust}$. The red circles are SPT DSFGs. The low-z and high-z samples are compiled by \cite{Gullberg2015} (see references therein). SPT0346-52 (red star) lies in the region surrounded by AGN-dominated galaxies. This diagnostic provided the motivation to search for X-ray signatures of an AGN in this extreme source.} 
\label{fig:LciiLfir_Td}
\end{figure}

The dust temperature of SPT0346-52, 52.4 $\pm$ 2.2 K \citep{Gullberg2015}, is higher than that of typical DSFGs and reaches into the territory of AGN-dominated sources (Figure \ref{fig:LciiLfir_Td}). SPT0346-52 has an $L_{\rm [CII]}$/$L_{\rm FIR}$ ratio consistent with FIR-luminous quasars at $z$ $\sim$ 6 and also shows an $L_{\rm [CII]}$ deficit relative to $L_{\rm FIR}$ and $L_{\rm CO(1-0)}$, which has been observed in AGN-dominated sources (\citealt{Stacey2010,Sargsyan2014,Gullberg2015}). It shows strong H$_2$O emission lines \citep{Weiss2013} similar to that of the strongly lensed quasars H1413+1143 and APM 08279+5255 \citep{Bradford2011}. SPT0346-52 is also optically obscured and does not show any indications of type-1 or type-2 AGN in deep VLT optical spectroscopy \citep{Hezaveh2013}.

DSFGs are in a unique phase of galaxy formation and evolution where the assembly of the stellar and supermassive black hole (SMBH) masses are believed to be closely coupled \citep{Alexander2012}. To test if \mbox{SPT0346-52} hosts an AGN and determine whether it is star formation-dominated or AGN-dominated, we resort to {\it Chandra}. Hard X-ray emission (rest-frame energies $>$ 2 keV) is the best indicator of AGN activity. These high-energy photons can penetrate through heavy obscuration, revealing the signature of the accreting black hole. A significant fraction of X-ray detected DSFGs have been found to be AGN-dominated in the X-ray, while some are powered by pure star formation (e.g., \citealt{Laird2010,Georgantopoulos2011,Johnson2013}). X-ray observations of the well-studied DSFG samples from the ALMA LABOCA E-CDF-S Submillimeter Survey (ALESS; \citealt{Wang2013}) reveal that 17\% of DSFGs appear to host an AGN. We here compare SPT0346-52 with these DSFGs and starbursts and quasars in the literature to understand the nature of the most extraordinary source found so far in the SPT survey. 

In addition to X-ray, radio also can be used to distinguish star-forming galaxies from AGN. Radio continuum emission from galaxies arises due to both thermal and non-thermal processes in massive star formation. These same massive stars also provide the primary sources of dust heating in the interstellar medium and the FIR emission is primarily due to the re-emitted starlight by dust. Star-forming galaxies that are not radio-loud AGN are observed to follow a tight FIR-to-radio correlation that holds over five orders of magnitude in galaxy luminosity (e.g., \citealt{Yun2001}). In contrast, radio-loud AGN will exhibit elevated radio emission above this relation (e.g., \citealt{Yun2001, Condon2002}). We utilize the Australia Telescope Compact Array (ATCA) to probe the radio continuum emission of SPT0346-52 to examine whether or not it is consistent with the FIR-to-radio correlation.

In this paper, we present the results from X-ray observations with the {\it Chandra} Observatory and radio continuum observations with ATCA to constrain the AGN activity, in combination with our existing multi-wavelength data. The Galactic column density towards \mbox{SPT0346-52} is $N_H$ =1.8 $\times$ 10$^{20}$ cm$^{-2}$.  We assume a  $\Lambda$CDM cosmology with $H_0$ =69.3 km s$^{-1}$ Mpc$^{-1}$, $\Omega_{m}$ = 0.286, and $\Omega_{\Lambda}$ = 0.713 (WMAP9; \citealt{Hinshaw2013}). We adopt the definition of $L_{\rm IR}$ to be integrated over rest-frame 8-1000 $\micron$ and $L_{\rm FIR}$ integrated over rest-frame 42.5-122.5 $\micron$. We assume a \cite{Chabrier2003} initial mass function (IMF) throughout the paper.

\section{Observations} \label{sec:data}

We present the new {\it Chandra} X-ray and ATCA radio continuum data for SPT0346-52 in this section. Previous near-IR to far-IR photometric data are summarized in Table \ref{tab:summary}.

\begin{table}
\centering
\caption{Multi-wavelength observed  flux densities in mJy.}
\begin{tabular}{lcc}
\hline
\hline
Telescope    & Wavelength &  SPT-S J034640-5204.9\\
\hline
 {\it HST}/WFC3 & 1.1$\mu$m	      &  $< 3.8 \times 10^{-4}$ \\
{\it HST}/WFC3  &1.6$\mu$m	      &  $<9.1 \times 10^{-4} $ \\
{\it Spitzer}/IRAC &3.6$\mu$m	      &  $<$ 0.0024                  \\
{\it Spitzer}/IRAC &4.5$\mu$m	      &  $<$ 0.0036                  \\
{\it Herschel}/PACS&100$\mu$m	      &  $<6$                            \\
{\it Herschel}/PACS&160$\mu$m	      &  $33 \pm 9$		    \\
{\it Herschel}/SPIRE &250$\mu$m    &  $122 \pm 11$              \\
{\it Herschel}/SPIRE &350$\mu$m    &  $181 \pm 14$              \\
{\it Herschel}/SPIRE &500$\mu$m    &  $204 \pm 15$              \\
APEX/LABOCA &870$\mu$m 	      &  $131 \pm 8$               \\
SPT &1.4mm                                     &  $46.0 \pm 6.8$            \\
SPT & 2.0mm                                    &  $11.6 \pm 1.3$            \\
ATCA &5.5cm                                   &  $<$ 0.114                   \\
ATCA &14.3cm                                 &  $<$ 0.213                    \\
\hline
\end{tabular}
\tablecomments{For the non-detections, the flux upper limits are given at 3$\sigma$. To derive the intrinsic flux densities, we divide the observed values by $\mu$ = 5.6 $\pm$ 0.1.}
\label{tab:summary}
\end{table}

\subsection{{\it Chandra} X-ray data}

SPT0346-52 was observed with the Advanced CCD Imaging Spectrometer (ACIS; \citealt{Garmire2003}) on board {\it Chandra} on 2015 July 29.  The source was placed at the aim point of the back-illuminated ACIS-S3 chip. The data were taken in Very Faint mode and were initially processed by the {\it Chandra} X-ray Center (CXC) using software version 10.4.1 and CalDB version 4.6.8. 

We reprocessed the data with the {\it Chandra} Interactive Analysis of Observations (CIAO; version 4.7) tool $chandra\_repro$.  All the bad pixels were removed and the standard grade (0,2,3,4,6), status, and good-time filters were applied. The net exposure time for the observation is 49.52 ks. We performed energy filtering on events into three {\it Chandra} bands: the soft (SB; 0.5-2.0 keV), hard (HB; 2-8 keV), and full (FB; 0.5-8.0 keV) bands. The soft and hard bands probe rest-frame energies 3.3-13.3 keV and 13.3-53.2 keV for $z$ = 5.656, respectively. No detectable X-ray emission is expected from the foreground lens. Existing optical and radio data show no evidence for an AGN in the lens. Moreover, the foreground lens is an elliptical galaxy (based upon the light profile fitting by \citealt{Ma2015}), and thus should also have negligible X-ray emission from star formation (more than three orders of magnitude below the detection threshold). 

\begin{figure*}
\centering
{\includegraphics[width=6.08cm,height=6.08cm]{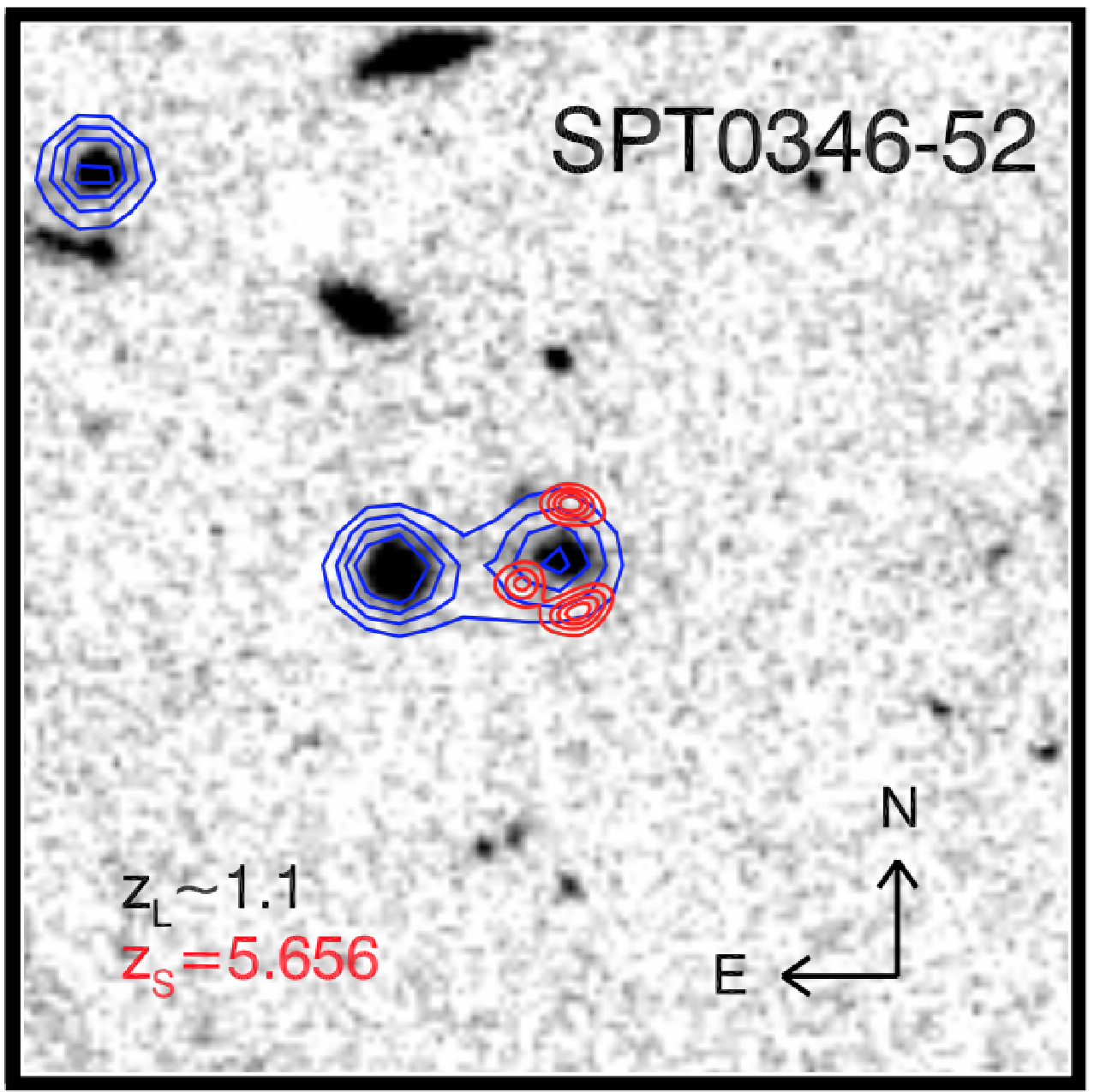}
 \includegraphics[width=6.06cm,height=6.06cm]{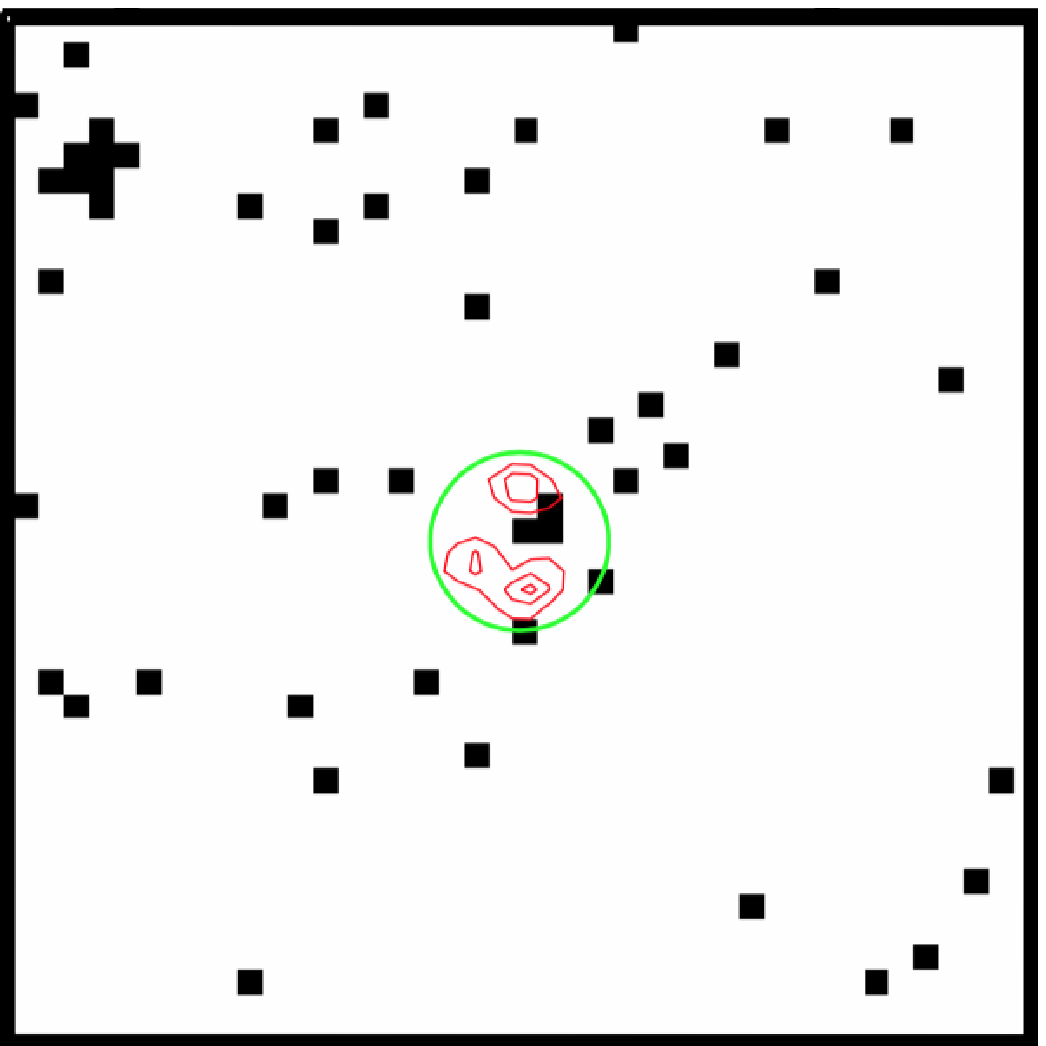}}  
\caption{$Left:$ 20$\arcsec$ $\times$ 20$\arcsec$ cutouts of SPT0346-52 showing the HST/WFC3 (gray), Spitzer/IRAC (blue contours), and ALMA band 7 (red contours) data. $Right:$ {\it Chandra} 0.5-8 keV full-band data. The green circle shows the source extraction aperture enclosing the ALMA contours (red). The energies of the three photons are 1.284 (upper right), 2.585 (lower left), and 2.009 (lower right) keV.}
\label{fig:data}
\end{figure*}

We matched the source to the position of ALMA (Figure \ref{fig:data}) and used a source extraction radius of 1.75$\arcsec$ enclosing all the lensed images, which is $\sim$1.3 times the 90\% encircled-energy aperture radius (at 0.3$\arcmin$ off-axis angle). The aperture was chosen such that it is large enough to enclose the ALMA contours without including too much background. The background counts were estimated by placing 78 circular apertures with the same size at random positions in the field.  We only detected 3.19 net source counts in the full band (0.73 in the SB and 2.47 in the HB).  Due to the low count level, we utilize the tool $aprates$ in the $srcflux$ script in CIAO to place a proper upper limit on the X-ray flux. This tool employs Bayesian statistics to compute the background-marginalized posterior probability distribution for source counts/flux. The posterior distribution can be used to determine flux value and confidence intervals or upper limits. The resultant FB flux is consistent with a non-detection with a 3 $\sigma$ upper limit of 6.0 $\times$ 10$^{-15}$ \flux.  We derive the rest-frame 0.5-8 keV apparent luminosity $L_{0.5-8 {\rm  keV}}$ (without absorption correction) using the following equation, 

\begin{equation}
L_{0.5-8 {\rm  keV}} = 4\pi d^2_{\rm L} f_{0.5-8 {\rm  keV}} (1+z)^{\Gamma_{\rm eff} - 2}
\label{eq:f2L}
\end{equation}

\noindent where $d_{\rm L}$ is the luminosity distance at $z$=5.656 and $\Gamma$ is the effective power-law photon index. In principle, the photon index can be derived from the hardness ratio, which is the ratio of the photon count rates in the HB and the SB. However we cannot derive a reliable hardness ratio based on the upper limits in both bands.  Instead, $\Gamma_{\rm eff}$ is fixed to 1.4 following \cite{Xue2011} and \cite{Wang2013}.

We then derive the rest-frame 0.5-8 keV absorption-corrected luminosity $L_{0.5-8 {\rm  keV, unabs}}$ by replacing $\Gamma_{\rm eff}$ with intrinsic photon index $\Gamma_{\rm int}$ and $f_{0.5-8 {\rm  keV}}$ with unabsorbed flux $f_{0.5-8 {\rm  keV, unabs}}$ in Equation \ref{eq:f2L}. We assume $\Gamma_{\rm int}$ = 1.8, a typical value for AGNs. The unabsorbed flux is estimated using the tool $modelflux$ within the $srcflux$ script.  We run simulations adopting Sherpa \citep{Freeman2001} models $xspowerlaw$$\times$$xszphabs$$\times$$xsphabs$  with fixed $\Gamma$ = 1.8 for the power-law model and hydrogen column density $N_{\rm H}$ for the (intrinsic and Galactic) absorption models. We scale the measured 3$\sigma$ upper limit on the flux by a fractional correction for absorption based on the typical intrinsic $N_{\rm H}$ (2.3 $\times$ 10$^{23}$ cm$^{-2}$) from the ALESS SMG sample \citep{Wang2013}. The absorption-corrected 3 $\sigma$ upper limit is $f_{0.5-8 {\rm  keV, unabs}}$ $<$ 7.6 $\times$ 10$^{-15}$ \flux. Since SPT0346-52 is gravitationally lensed, we further correct the X-ray flux and luminosity for lensing magnification assuming there is no differential magnification between the FIR (i.e., ALMA) and the X-ray emission \citep{Hezaveh2012}.  The magnification-corrected FB flux and luminosity upper limits are listed in \mbox{Table \ref{tab:chandra}}.

\begin{table*}
\centering
\caption{{\it Chandra} X-ray properties of SPT0346-52}
\begin{tabular}{lcccccccc}
\hline
\hline
Source Name    &  Redshift & Exptime &  Full-band & Background & \multicolumn{2}{c}{Full-band Flux (3$\sigma$) }                          &  \multicolumn{2}{c}{Full-band Luminosity (3$\sigma$) }             \\
                          &                &       (ks)     &  (count)                & (count)   & \multicolumn{2}{c}{ ($\times$ 10$^{-15}$ \flux) }                       &           \multicolumn{2}{c}{($\times$ 10$^{44}$ \lum)}      \\
                          &                &                  &                  &  &$f_{0.5-8 {\rm  keV}}$      &    $f_{0.5-8 {\rm  keV, unabs}}$  &  $L_{0.5-8 {\rm  keV}}$ &   $L_{0.5-8 {\rm  keV, unabs}}$   \\
\hline
SPT0346-52 &	5.656  & 49.52& 3.19 & 0.81     &  $<$ 1.07              &      $<$ 1.36        &  $<$ 1.20  &  $<$  3.23                      \\
\hline
\end{tabular}
\label{tab:chandra}
\end{table*}

\begin{figure*}
\centering
{\includegraphics[width=18.5cm, height=16.2cm]{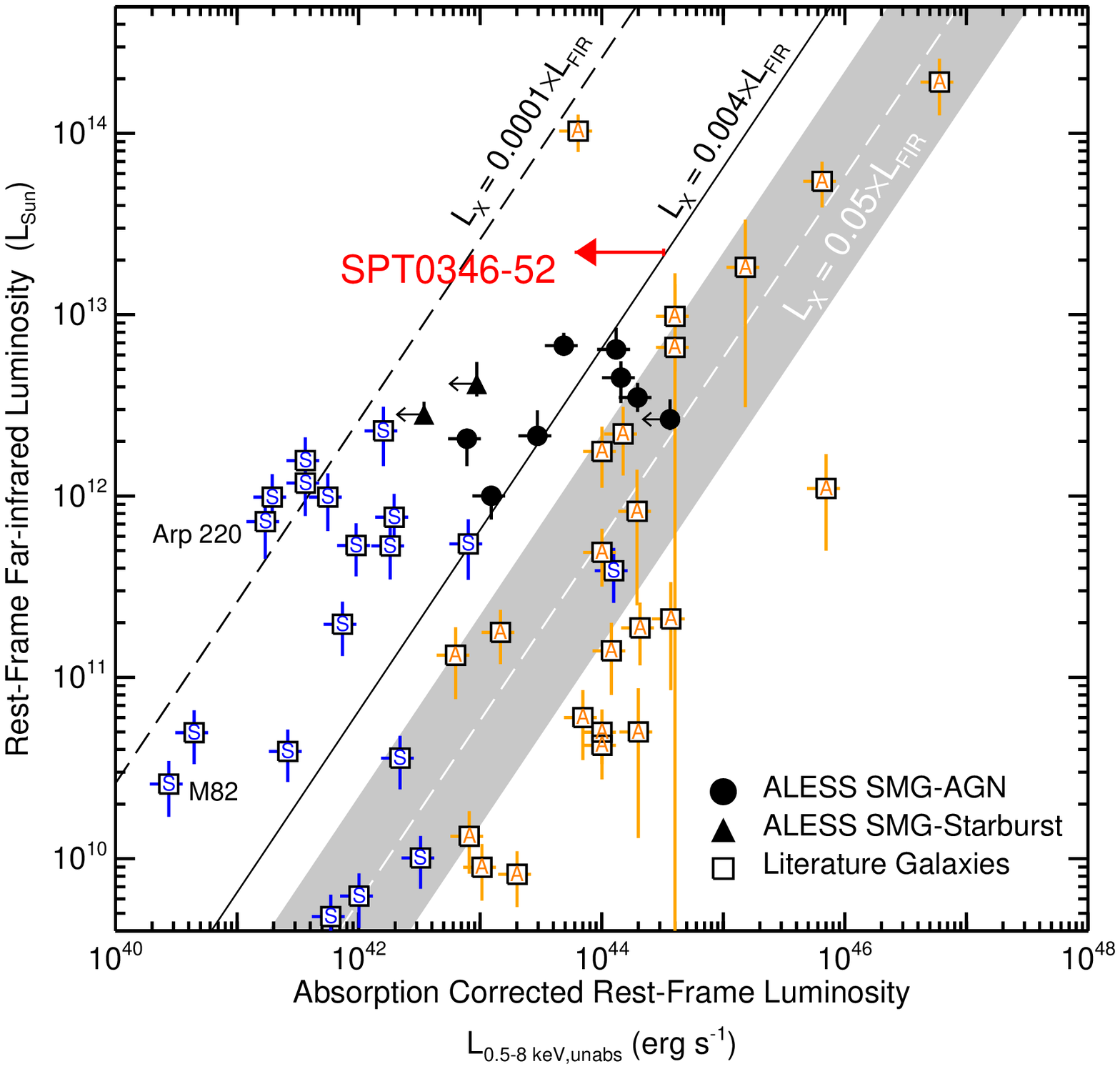}}  
\caption{$L_{\rm FIR}$ vs. $L_{\rm X}$ for SMGs and literature galaxies. SPT0346-52 is the red arrow representing the 3 $\sigma$ upper limit on the absorption-corrected $L_{\rm X}$. We have corrected $L_{\rm FIR}$ and $L_{\rm X}$ for lensing magnification. We show the ALESS SMGs from \cite{Wang2013}. The X-ray detected SMGs are marked as starburst-classified SMGs and AGN-classified SMGs. The dividing line between star formation-dominated and AGN-dominated appears to be $L_{\rm X}$ = 0.004 $\times$ $L_{\rm FIR}$, which is the median ratio of SCUBA SMGs found by  \cite{Alexander2005a}. The squares are literature galaxies that are classified as star formation-dominated (labeled with blue `S') or AGN-dominated (labeled with orange `A'). These data points are compiled by \cite{Alexander2005a} (see references therein) and we have re-fit the FIR photometry in a consistent manner as described in \cite{Gullberg2015}. We assume 30\% error on $L_{\rm X}$ for the literature galaxies. The dashed white line and the gray region represent the fiducial $L_{\rm FIR}$ vs. $L_{\rm X}$ relation and its standard deviation for the quasars in \cite{Elvis1994}. The $L_{\rm X}$-to-$L_{\rm FIR}$ ratio for pure starbursting systems (black dashed line) is about 2 orders of magnitude lower.}
\label{fig:LFIR_Lx}
\end{figure*}

\subsection{ATCA radio data}

 SPT0346-52 was observed with ATCA for 3960s at 5.5 and 9.0 GHz and 4068s at 2.1 GHz on 2012 January 25 in the 6A array configuration using the CABB in the 1M-0.5k mode. The data was reduced in the same manner as in \cite{Aravena2013}. The resultant synthesized beam sizes are  7.7$\arcsec$ $\times$ 5.2$\arcsec$ at 2.1 GHz, 3.3$\arcsec$ $\times$ 2.2$\arcsec$ at 5.5 GHz and 2.1$\arcsec$ $\times$ 1.3$\arcsec$ at 9.0 GHz. The continuum was not detected in any band and we place 3 $\sigma$ upper limits (i.e., 3 $\times$ rms noise values calculated within a 1$\arcmin$ region around the source position) of 0.213 mJy at 2.1 GHz, 0.114 mJy at 5.5 GHz and 0.138 mJy at 9.0 GHz on the radio emission from the source.

\section{Discussion}

\subsection{Observational/Empirical View}
\label{sec:Lx_Lfir}

We place SPT0346-52 on the $L_{\rm FIR}$-$L_{\rm X}$ plane (\mbox{Figure \ref{fig:LFIR_Lx}}) in the context of X-ray quasars and starburst galaxies. There, we compare it with other SMGs that are identified as SMG-AGN or SMG-starbursts to determine if it is star formation-dominated or AGN-dominated. As shown in Figure \ref{fig:LFIR_Lx}, the starburst galaxies (squares labeled with `S') from the literature occupy different locations than AGN-dominated galaxies (squares labeled with `A') and quasars (dashed line and the gray region).  The well-studied unobscured quasars of \cite{Elvis1994} provide the fiducial X-ray to FIR luminosity ratios (the median ratio is $L_{\rm X}$/$L_{\rm FIR}$ = 0.05 and the gray region indicates the standard deviation) for AGN-dominated sources. The luminosity ratio for pure starburst galaxies is about two orders of magnitude lower. 

For SMGs, which may involve the co-evolution of super massive black holes and the host galaxies, a heterogeneous population has been observed.  The dividing line between starbursts and AGN is typically taken to be $L_{\rm X}$/$L_{\rm FIR}$ $\sim$ 0.004 \citep{Alexander2005a}. The extensively studied ALESS SMGs from \cite{Wang2013} are consistent with this notion.  The X-ray to FIR ratio upper limit for SPT0346-52 is 0.0038, slightly leftward of the dividing line, which indicates that it is consistent with being starburst-dominated in the X-ray.  One caveat is that if the absorption for SPT0346-52 is exceptionally high, the X-ray upper limit could be weakened. For example, if $N_{\rm H}$ is 4 times higher (the highest $N_{\rm H}$ for ALESS SMGs) than what we adopted, the upper limit will be 1.3 times higher, which would move the limit slightly to the right of the dividing line between starbursts and AGN in Figure \ref{fig:LFIR_Lx}. For this reason, we also cannot exclude Compton thick AGN by $L_{\rm FIR}$-$L_{\rm X}$ alone (e.g., \citealt{Murphy2009}). The second potential caveat is AGN X-ray variability. Using the relation between X-ray luminosity and variability found by \cite{Lanzuisi2014}, the fractional variability of \mbox{SPT0346-52} is expected to be 30\%. A third potential caveat is the possibility of differential magnification between the star formation region and the AGN (e.g., \citealt{Hezaveh2012}). We have assumed a constant magnification of $\mu$ = 5.6. It could be that the star formation region is more magnified than the AGN. The maximum difference in magnification between components of this galaxy found by \cite{Spilker2015} is $\Delta\mu$ $\sim$ 2, which would move the red upper limit in Figure \ref{fig:LFIR_Lx} by the same amount. However, without a robust X-ray detection and resolved X-ray image of the system, it is impossible to say any more. 

If we adopt $L_{\rm X}$/$L_{\rm FIR}$ = 0.05 for typical quasars and take the ratio ($L_{\rm X}$/$L_{\rm FIR}$)$_{\rm SPT0346-52}$/($L_{\rm X}$/$L_{\rm FIR}$)$_{\rm quasars}$ following \cite{Alexander2005a}, the AGN fractional contribution to the FIR luminosity is estimated to be at most 8\%. We note that an AGN in SPT0346-52 would be fainter than the quasars studied by \cite{Elvis1994}, and that Seyfert 1 galaxies tend to be relatively more X-ray luminous than quasars. Using the extensive Seyfert observations of \cite{Rush1996a} would shift the grey band in Figure \ref{fig:LFIR_Lx} (and AGN/starburst boundary) to the right by a factor of a few.  This would mean that the FIR contribution of any AGN in \mbox{SPT0346-52} would be even smaller than our upper limit of 8\%. Although we cannot completely rule out the presence of an AGN in the system, it is certainly the star-formation that is dominating the FIR emission, which is confirmed by fitting spectral energy distributions including an AGN component in Section \ref{sec:SEDfitting}.

\subsection{Constraining AGN fraction through SED fitting}
\label{sec:SEDfitting}

We previously performed SED fitting on \mbox{SPT0346-52} with CIGALE FORTRAN assuming no AGN contribution in the IR \citep{Ma2015}, which is consistent with the NIR photometric upper limits and FIR detections. Now we employ CIGALE PYTHON \citep{Roehlly2014}, which includes up-to-date star formation history models, stellar population synthesis models, IR re-emission models, and AGN templates, to constrain the potential AGN contribution in the IR. The photometric data points used in the SED fitting are listed in Table \ref{tab:summary}. We adopt a \cite{Chabrier2003} IMF and the \cite{Bruzual2003} stellar population synthesis models.  For star formation history, we assume a delayed-$\tau$ model which rises at early ages and then declines exponentially. This form of star formation history is generally expected for especially high-redshift galaxies (\citealt{Pacifici2013,Simha2014,daCunha2015}). We use a combination of \cite{Dale2014} IR models accounting for the dust emission from the stellar component and \cite{Fritz2006} AGN templates.  The \cite{Fritz2006} AGN models take into account two emission components associated with the AGN: a power-law from the central source and the thermal and scattering dust-torus emission. The relative normalization of these components is handled through a parameter, fracAGN, which is the fractional contribution of the AGN to the total IR luminosity ($L_{\rm IR, total} = L_{\rm Starburst} + L_{\rm AGN}$). The model curve is also extended to radio wavelengths. The extension relies on the well-established FIR-to-radio correlation for star-forming galaxies. CIGALE does not include synchrotron emission from AGN. 

\begin{figure}
\centering
{\includegraphics[width=9cm, height=7.3cm]{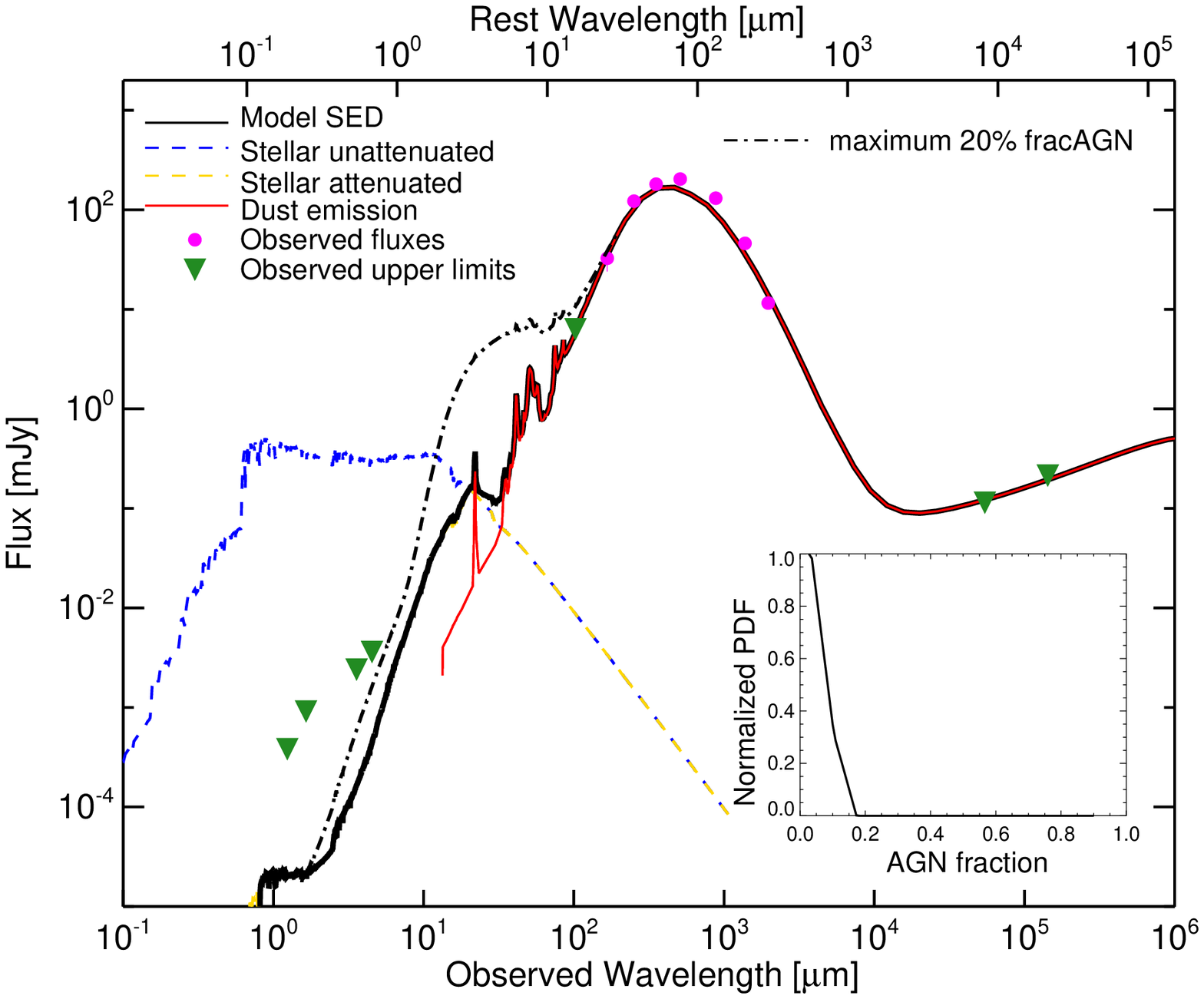}}    
\caption{The best-fit SED (black) from CIGALE PYTHON. Different components are color-coded. The data points from left to right are {\it HST}/WFC3 F110W+F160W, {\it Spitzer}/IRAC 3.6 $\micron$ + 4.5 $\micron$, {\it Herschel}/PACS 100 $\micron$ + 160 $\micron$, {\it Herschel}/SPIRE 250 $\micron$ + 350 $\micron$ + 500 $\micron$, APEX/LABOCA 870 $\micron$, SPT 1.4 mm + 2.0 mm, and ATCA  5.5 GHz + 2.1 GHz. We do not include the ATCA 9.0 GHz upper limit, which is less constraining, because the source may be resolved at this frequency. The inset plot shows the normalized probability distribution function of the AGN fraction in the total IR luminosity. The AGN most probably contributes less than 5\%. The dash-dotted line is the model curve with the maximum 20\% AGN contribution to the IR. In this model, the AGN contribution does not extend to the radio part of the SED.}
\label{fig:sed}
\end{figure}

A pure starburst SED remains the best-fit SED (i.e., with the minimal reduced $\chi^2$), constrained by all the photometric detections and upper limits from NIR to radio wavelengths (Figure \ref{fig:sed}).  Parameters are analyzed under a Bayesian approach generating a posterior probability distribution function. The SED fitting failed to tightly constrain fracAGN given the lack of mid-IR photometric points (which differentiate between AGN and star-forming galaxies) and the loose constraint from the NIR.  The posterior probability distribution suggests that the AGN component contributes at most $\sim$20\% to the total IR luminosity, with the highest probability being assigned to a contribution of 0 - 5\%. This is consistent with the estimation from the $L_{\rm X}$/$L_{\rm FIR}$ ratio in Section \ref{sec:Lx_Lfir}. The model SED with the maximum 20\% AGN fraction is also plotted in Figure \ref{fig:sed}, which is inconsistent with the 100 $\micron$ Herschel/PACS upper limit. 

The ATCA radio continuum upper limits at 2.1 GHz and 5.5 GHz are consistent with the FIR-to-radio correlation for star-forming galaxies. The radio part of the SED for radio-loud AGN would be at least a factor of $\sim$ 2 higher (e.g., \citealt{Rush1996a,Rush1996b,Moric2010}).

\begin{figure}
\centering
{\includegraphics[width=10cm, height=7.5cm]{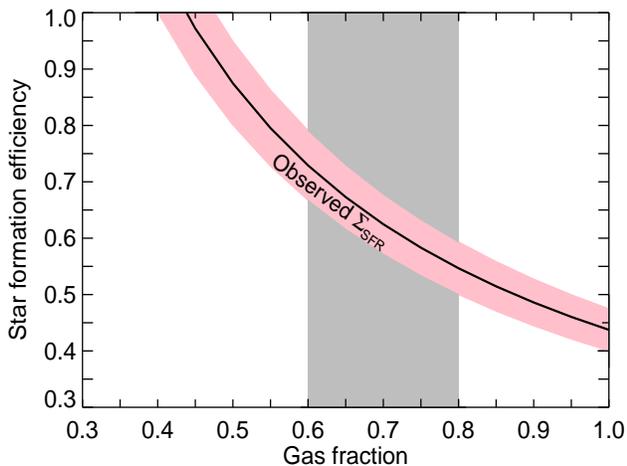}}  
\caption{Theoretical expectations based on ``maximum starbursts" for the star formation efficiency and gas fraction of SPT0346-52. Very high gas fraction and star formation efficiency are required to explain the observed $\Sigma_{\rm SFR}$ (the solid line and pink band). The gray band shows the average gas fractions of SPT DSFGs derived from CI observations found by Bothwell et al. in preparation. }
\label{fig:sfe}
\end{figure}

\subsection{Theoretical Expectations}

We have provided evidence that SPT0346-52 is star formation-dominated in the IR. Since our SFR is mainly constrained by the FIR photometric data, $L_{\rm IR}$ being almost all from star formation suggests that the inferred SFR reflects the true SFR.  We examine whether this observed high SFR can be physically explained in the framework of ``maximum" starbursts \citep{Elmegreen1999} where a substantial fraction $\epsilon$ of available gas is consumed to make stars. Following \cite{Tacconi2006}, the maximum (``Schmidt-law") SFR can be written as 
\begin{equation}
SFR_{\rm max} =  \frac{\epsilon f_g M_{\rm tot}}{t_{\rm dyn}} = 630\Big(\frac{\epsilon}{0.1}\Big) \Big(\frac{f_g}{0.4}\Big) \Big(\frac{v_c}{400}\Big)^3    (M_{\sun} {\rm yr^{-1}})
\end{equation}

\noindent where \cite{Elmegreen1999} defines $\epsilon$ as the star formation efficiency, $f_g$ is the gas fraction, $M_{\rm tot}$ is the total (dynamical) mass of the system, $t_{\rm dyn}$ is the dynamical time, and $v_c$ is the circular velocity in km s$^{-1}$.  We use the definition in \cite{Tacconi2006},  $v_c$ = 0.67$\times$$v_{\rm FWHM}$ =  410 km s$^{-1}$ ($v_{\rm FWHM}$ = 613 $\pm$ 30 km s$^{-1}$ measured from the low-J CO(2-1) line; \citealt{Aravena2016}).  Assuming a gas fraction of 0.3 - 0.8, we utilize the observed SFR of 3600 $\pm$ 300 $M_{\sun} {\rm yr^{-1}}$ \footnote{Here we adopt the SFR converted directly from $L_{\rm IR}$ using the \cite{Kennicutt1998} conversion factor assuming a \cite{Chabrier2003} IMF where SFR = 1.0 $\times$ 10$^{-10}$ $L_{\rm IR}$ ($L_{\sun}$).} to derive $\epsilon$, which turns out to be in the range of 0.3-0.7. This range is higher than the typical star formation efficiency (0.15-0.2) observed in DSFGs \citep{Tacconi2006}.

\begin{figure}
\centering
{\includegraphics[width=9cm, height=6.7cm]{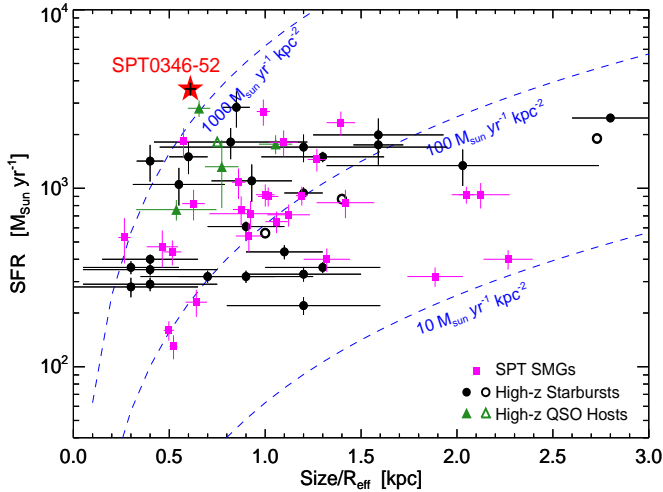}}    
\caption{Star formation as a function of dust continuum size for high-redshift dusty starburst galaxies, illustrating that SPT0346-52 (red star with the cross error bar) has an exceptionally large SFR surface density. The dashed lines show constant $\Sigma_{\rm SFR}$ values. The magenta squares are SPT SMGs from \cite{Spilker2016}. The black circles are literature starburst galaxies at $z$ $>$ 3 with dust continuum size measurements. The green triangles are quasar host galaxies. The open circles and triangles do not have reported error bars. The literature galaxies are drawn from \cite{Younger2008}, \cite{Walter2009}, \cite{Magdis2011}, \cite{Walter2012}, \cite{Fu2012}, \cite{Bussmann2013}, \cite{Carniani2013}, \cite{WangR2013}, \cite{Cooray2014}, \cite{DeBreuck2014}, \cite{Riechers2014}, \cite{Yun2015}, \cite{Simpson2015}, and \cite{Ikarashi2015}. SFR is based upon $L_{\rm IR}$. $L_{\rm FIR}$ is converted to $L_{\rm IR}$ by multiplying 1.65 when necessary. }  
\label{fig:size}
\end{figure}

We further examine the SFR surface density $\Sigma_{\rm SFR_{max}}$ expected from this framework. 

\begin{equation}
\begin{aligned}
\Sigma_{\rm SFR_{max}} & =  \frac{SFR_{\rm max}}{\pi R^2} \\
                                         & = 43 \Big(\frac{\epsilon}{0.1}\Big) \Big(\frac{f_g}{0.4}\Big) \Big(\frac{\Sigma_{\rm tot}}{5000}\Big) \Big(\frac{R}{2}\Big)^{-1}   (M_{\sun} {\rm yr^{-1}} {\rm kpc^{-2}})
\end{aligned}
\end{equation}

\noindent where $\Sigma_{\rm tot}$ is the total (dynamical) mass density within radius $R$, in units of $M_{\sun}$ pc$^{-2}$. For SPT0346-52, $R$ = 1.8 $\pm$ 0.2 kpc and $M_{\rm tot}$ = (1.5 $\pm$ 0.2) $\times$ 10$^{11}$ $M_{\sun}$, derived in a spatially resolved CO imaging study by \cite{Spilker2015}. The observed $\Sigma_{\rm SFR}$ = 1540 $\pm$ 130 $M_{\sun} {\rm yr^{-1}} {\rm kpc^{-2}}$ \footnote{To derive the SFR surface density, we divide the SFR by \mbox{2$\pi$R$_{\rm eff}^2$} where the factor of 2 corresponds to the half-light radius R$_{\rm eff}$ from the dust emission.}, in which the dust emission (i.e., the stellar emission reprocessed by dust) is distributed in a compact area with an effective radius of 0.61 $\pm$ 0.03 kpc, surpasses the theoretical $\Sigma_{\rm SFR_{max}}$ by at least a factor of $\sim$ 2. It is one of the highest star formation densities of any known galaxy in the Universe (\citealt{Rujopakarn2011,Diamond-Stanic2012}), although the nuclei of Arp 220 have $\Sigma_{\rm SFR}$ $\sim$10$^4$ $M_{\sun} {\rm yr^{-1}} {\rm kpc^{-2}}$ \citep{Barcos2015}. Figure \ref{fig:size} shows how SPT0346-52 compares in SFR surface density to other starburst galaxies (black) and starbursts found in quasar host galaxies (green) at $z$ $>$ 3 in the literature. SPT0346-52 has an order of magnitude higher SFR than most other sources that lie within a factor of a few in star formation surface density. SPT0346-52  stands out as the most extreme source at high redshift.

If the starburst is supported by radiation pressure on dust grains in a disk, $L_{\rm IR}$ is consistent with the Eddington-limited luminosity but the SFR surface density is above the $\sim$ 1000 $M_{\sun} {\rm yr^{-1}} {\rm kpc^{-2}}$ theoretical limit by \cite{Thompson2005}. The Eddington limit depends upon opacity, and the observed SFR for SPT0346-52 lies in between the SFR$_{\rm max, thin}$ for the optically thin limit and SFR$_{\rm max, thick}$ for the optically thick limit from \cite{Younger2008}. The observed brightness temperature ($\sim$ 50 K) is comparable to the dust temperature, which suggests that the gas may be approaching the optically thick regime (\citealt{Murray2005, Younger2008}). Thus, the star formation activity in this source, while very vigorous, may still be sub-Eddington. It is natural to suspect that the radiation pressure would drive outflows of cold dusty gas, which have been commonly observed from starbursting galaxies including ultra-luminous IR galaxies (e.g., \citealt{Martin2005,Spoon2013,Veilleux2013}).

\subsection{Possible explanations for the $\Sigma_{\rm SFR}$}

Figure \ref{fig:sfe} demonstrates the allowed $\epsilon$ and $f_{gas}$ in the theoretical framework of ``maximum starbursts". This observed extremely high $\Sigma_{\rm SFR}$ may be explained by an especially high star formation efficiency ($\epsilon$ $>$ 0.4) relative to what has been observed in DSFGs. The gas fraction is constrained to be at least 40\%. So far we have only detected a handful of DSFGs at $z$ $>$ 5, only a few hundred million years from the Big Bang (\citealt{Capak2011, Combes2012, Walter2012, Riechers2013, Strandet2016}). These sources are expected to harbor larger gas reservoirs available for star formation and be able to sustain a more elevated star formation efficiency than typical DSFGs \citep{Bethermin2015}. Gas fractions of SPT DSFGs derived using low-J CO observations are in the range of 0.3 - 0.8 \citep{Aravena2016}. Bothwell et al. in preparation found that SPT DSFGs (SPT0346-52 is not in this sample) on average have very high gas fractions ($f_{gas}$ $\sim$ 0.6 - 0.8) based on atomic carbon observations. The very high gas fractions could raise the $\Sigma_{\rm SFR}$ without invoking extremely high star formation efficiency.  

Emission from SPT0346-52 has proved to be beyond the reach of our existing {\it HST} and {\it Spitzer} data \citep{Ma2015}, which has prevented a detailed characterization of the established stellar mass. Thus, SPT0346-52 would be an ideal object for follow-up observations with JWST. 

In this paper we presented a pilot X-ray observation of a single extreme star-forming galaxy at high redshift with {\it Chandra}. Studying a sample of such vigorously star forming galaxies in the early universe will help us constrain the formation and co-evolution of the massive galaxies and supermassive black holes.

\section*{acknowledgments}

We thank the anonymous referee for insightful and constructive comments which have significantly improved the paper. The scientific results reported in this article are based on observations made by the {\it Chandra} X-ray Observatory and the Australia Telescope Compact Array. This research has made use of software provided by the {\it Chandra} X-ray Center (CXC) in the application packages CIAO and Sherpa. The Australia Telescope Compact Array is part of the Australia Telescope National Facility which is funded by the Australian Government for operation as a National Facility managed by CSIRO. The associated {\it HST},  {\it Spitzer}, and ALMA data are from PID12659, PID10094, and PID2011.0.00958.S, respectively. 

We acknowledge support from the {\it Chandra} grant GO5-16116A and the U.S. National Science Foundation under grant No. AST-1312950. M.A. acknowledges partial support from FONDECYT through grant 1140099. Y.H. acknowledges support from Hubble Fellowship grant 51358.001-A awarded by the Space Telescope Science Institute. This research has made use of NASA's Astrophysics Data System.

\end{document}